\documentclass{article}

\usepackage{PRIMEarxiv}

\usepackage[utf8]{inputenc} 
\usepackage[T1]{fontenc}    
\usepackage{hyperref}       
\usepackage{url}            
\usepackage{booktabs}       
\usepackage{amsfonts}       
\usepackage{nicefrac}       
\usepackage{microtype}      
\usepackage{lipsum}
\usepackage{fancyhdr}       
\usepackage{graphicx}       
\graphicspath{{media/}}     
\usepackage{tabularx}
\usepackage{amsmath} 
\usepackage{xcolor}

\title{VOLTRON: Detecting Unknown Malware Using Graph-Based Zero-Shot Learning
}

\author{
  M. Tahir Akdeniz \\
  WISE Lab, Hacettepe University \\
  Ankara, Turkey \\
  \texttt{muhammedtahirakdeniz@gmail.com}
  \And
  Zeynep Yesilkaya \\
  WISE Lab, Hacettepe University \\
  Ankara, Turkey \\
  \texttt{zeynpyesilkaya@gmail.com}
  \And
  I. Enes Kose \\
  WISE Lab, Hacettepe University \\
  Ankara, Turkey \\
  \texttt{enes0kose@gmail.com}
  \And
  I. Ulas Unal \\
  WISE Lab, Hacettepe University \\
  Ankara, Turkey \\
  \texttt{ismailulasunal@gmail.com}
  \And
  Sevil Sen \\
  WISE Lab, Hacettepe University \\
  Ankara, Turkey \\
  \texttt{ssen@cs.hacettepe.edu.tr}
}

\begin{document}
\maketitle

\begin{abstract}
The persistent threat of Android malware presents a serious
challenge to the security of millions of users globally. While many machine learning-based methods have been developed to detect these threats, their reliance on large labeled datasets
limits their effectiveness against emerging, previously unseen malware families, for which labeled data is scarce or nonexistent.

To address this challenge, we introduce a novel zero-shot
learning framework that combines Variational Graph Auto-Encoders (VGAE) with Siamese Neural Networks (SNN) to
identify malware without needing prior examples of specific
malware families. Our approach leverages graph-based representations of Android applications, enabling the model to
detect subtle structural differences between benign and malicious software, even in the absence of labeled data for new
threats.

Experimental results show that our method outperforms the state-of-the-art MaMaDroid, especially in zero-day malware detection. Our model achieves 96.24\% accuracy and 95.20\% recall for unknown malware families, highlighting its robustness against evolving Android threats.
\end{abstract}


\section{Introduction}
Malware is one of the most significant threats in cybersecurity today. With the number of mobile devices now surpassing that of desktop devices, attackers are increasingly focusing their efforts on targeting mobile platforms. Moreover, while generating malware traditionally required substantial technical skills, advancements in generative artificial intelligence (AI) have lowered the barrier for attackers, making it easier to create malicious software \cite{barrett2023identifying}. Additionally, large language models can produce different malware variants for similar prompts due to the inherent ambiguity in natural languages. Consequently, the threat has become more dangerous than ever. 

Considering the vast size and complexity of modern malware, machine learning has become an essential part of effective malware security solutions. Its ability to adapt and analyze large datasets in real-time makes it indispensable for detecting and mitigating sophisticated cyber threats. However, these proposals might have pitfalls in their design, implementation, and evaluation~\cite{arp2022and}. Moreover, they are often evaluated using known malware families included in the training data, which means their performance on new, unknown malware is typically not represented. When evaluated on malware from a different time frame than the training period, their performance can drop dramatically~\cite{pendlebury2019tesseract}. Therefore, this study particularly focuses on detecting unknown malware from new families using zero-shot learning techniques.

In the domain of Android malware detection, zero-shot learning is crucial for identifying zero-day malware~\cite{barros2022malware,cen2024zero}, new and previously unknown threats that existing security systems have not encountered. This learning paradigm involves classifying instances from classes not present in the training dataset, which is particularly important given the rapid evolution of malware that leads to the emergence of new families of threats. Zero-shot learning addresses the challenge where the training samples do not cover all the classes it aims to classify~\cite{deldar2023deep}. This approach is essential in scenarios where traditional deep learning models~\cite{liu2022deep}, which rely heavily on large-scale labeled datasets, struggle to adapt to the continuous emergence of novel malware families.

This study proposes a new graph-based approach utilizing zero-shot learning for detecting unknown malware. In this approach, each application is represented by sensitive API call graphs to capture and illustrate the semantic relationships between API calls. These graphs are then transformed into low-dimensional representations using a Variational Graph Auto-Encoder (VGAE) for malware detection. These representations are processed by a Siamese Neural Network (SNN) to predict similarity scores between application pairs. Finally, the combination of VGAE and SNN is used to assess the dissimilarity of new malicious software compared to benign applications within a zero-shot learning framework. Each component of this system contributes uniquely, akin to how individual lions combine to form the powerful robot VOLTRON in the animated series. Thus, our approach is named VOLTRON, reflecting the synergy of its elements to effectively detect unknown malware.

The main contributions of the proposed approach are summarized as follows:

\begin{itemize}
    \item VOLTRON is the first zero-shot learning approach proposed for detecting new Android malware, addressing the critical need given the rapid evolution of Android malware. This approach is highly significant in adapting to the continuously changing landscape of cyber threats.
    \item It is trained and evaluated on one of the largest datasets, KronoDroid~\cite{guerra2021kronodroid}, which includes 218 malware families and approximately 55,000 samples of both malware and benign software.
    \item VOLTRON's ability to detect unknown malware samples is thoroughly evaluated on unknown samples from 54 new malware families, achieving a detection rate of 95.20\%, compared to MaMaDroid's 89.11\%. A time-based analysis is also conducted to assess its detection efficiency for malware that emerges after the training period.
    \item The code for VOLTRON is shared to facilitate the reproducibility of our approach within the research community\footnote{\url{https://github.com/voltron-research-team/voltron}}.
\end{itemize}

The paper is organized as follows: Section~\ref{sec:relatedwork} presents a literature review on malware detection, emphasizing graph-based approaches for Android malware detection, as well as few-shot and zero-shot learning methods in this domain. Section~\ref{sec:methodology} details the proposed approach, including the steps involved, API call graph construction, Variational Graph Auto-Encoder (VGAE), Siamese Neural Network (SNN), and Zero-Shot Learning (ZSL). Section~\ref{sec:evaluation} introduces the dataset used for training and testing, outlines the experimental setup, and provides a detailed discussion of the evaluation results. Section~\ref{sec:limitations} addresses the limitations of the proposed approach, and Section~\ref{sec:conclusion} offers concluding remarks on the work.

\section{Related Work}
\label{sec:relatedwork}

Since the introduction of the first Android malware dataset, MalGenome~\cite{zhou2012dissecting}, numerous studies have focused on Android malware detection. In this section, we primarily focus on two recent and promising areas: graph-based approaches and few-shot learning techniques. Although these methods have been proposed separately for malware detection, our proposal aims to integrate them, leveraging the strengths of both to enhance detection accuracy, particularly in scenarios with limited labeled data.

Mariconti et al. developed MaMaDroid~\cite{onwuzurike2019mamadroid}, a static analysis-based system that constructs behavioral models by extracting call graphs from applications. It then abstracts these graphs into simplified API call sequences, such as packages or families, to manage complexity. These sequences are used to build a Markov-chain-based model that captures application behavior efficiently. Similarly, Hou et al.~\cite{hou2017hindroid} introduced HinDroid, which employs a heterogeneous information network (HIN) to model the relationships between apps and APIs, allowing for the capture of interactions that might be missed by simpler models.

Recent advancements have integrated deep learning techniques with API call sequence analysis. Cui et al. developed Api2Vec~\cite{cui2023api2vec}, while Zhang et al. introduced APIGraph~\cite{zhang2020enhancing}. These approaches address challenges such as API call interleaving and classifier aging, aiming to enhance the robustness of detection models and reduce the frequency of retraining, thereby maintaining performance as malware evolves.

With advancements in Graph Neural Networks (GNNs), such approaches have been proposed to improve Android malware detection. Gao et al.~\cite{gao2021gdroid} introduced GDroid, the first study utilizing Graph Convolutional Networks (GCNs) for malware classification. It maps applications and their API interactions into a heterogeneous graph. By reframing the detection task as a node classification problem, this approach identifies relationships between API calls and application behavior. When applied to well-known Android malware datasets (Drebin~\cite{arp2014drebin}, MalGenome~\cite{zhou2012dissecting}, and AMD~\cite{wei2017deep}), GDroid demonstrated a clear improvement in detection accuracy and robustness, highlighting the potential of GNNs to handle the intricate dependencies and interactions typical of Android malware.

John et al.~\cite{john2020graph} explored the use of GCNs with system call graphs, focusing on centrality measures to capture dependencies between system calls, even in the presence of obfuscated malware. Alam et al.~\cite{alam2017droidnative} introduced DroidNative, which detects native code malware by leveraging control flow patterns represented as graphs, aiming to address the impact of obfuscations. Lo et al.~\cite{lo2022graph} addressed the common issue of over-smoothing in deep GNNs by incorporating Jumping Knowledge techniques to enhance the analysis of function call graphs. Hei et al.~\cite{hei2021hawk} developed Hawk, which uses Heterogeneous Graph Attention Networks to model Android entities and their behavioral relationships within a heterogeneous information network (HIN), aiming to reduce training time.

In summary, these studies collectively highlight the shift towards advanced machine learning and graph-based techniques aimed at improving detection accuracy, model resilience, and adaptability. However, these approaches generally require large datasets and may struggle to perform effectively when only a small number of samples are available. Therefore, few-shot and/or zero-shot learning approaches are investigated to detect malware with limited labeled data.

While it is a relatively new area to explore, most proposed studies on few-shot malware detection focus on the Windows malware domain. Rong et al.~\cite{rong2021umvd} presented UMVD-FSL, which detects unseen malware variants by converting network traffic data into grayscale images and using a prototype-based model for classification. Tang et al.~\cite{tang2020convprotonet} introduced ConvProtoNet, proposing a convolutional induction module and utilizing meta-learning to adapt to new malware types without requiring fine-tuning. Liu et al.~\cite{liu2022deep} proposed FewM-HGCL, a self-supervised framework that models the execution behavior of each malware variant as a heterogeneous graph. FewM-HGCL employs heterogeneous graph contrastive learning to empower graph attention networks (GATs) in learning graph-level representations for few-shot malware variants. This approach utilizes graph instance-based discrimination and data augmentation techniques, such as API attribute masking, to generate samples for learning. By leveraging GATs to capture detailed dependencies and relationships within graphs, FewM-HGCL enhances the model's ability to generalize from a limited number of samples, which is essential for few-shot malware detection.

Zero-shot learning (ZSL) has also been explored as a strategy for detecting previously unseen Windows malware threats. Barros et al.~\cite{barros2022malware} introduced Malware-SMELL, a method that classifies malware using visual representation in a novel S-Space, which helps distinguish unseen malware. Zahoora et al.~\cite{zahoora2022zero} proposed DCAE-ZSL, a deep contractive autoencoder-based approach that uses ZSL to detect zero-day ransomware by extracting features from known and unknown ransomware and using a heterogeneous voting ensemble for inference.

There are only two notable few-shot-based studies on Android malware detection. Zhou et al.~\cite{zhou2024famcf} proposed a framework called FAMCF, which utilizes static analysis features such as permissions, API calls, and opcodes, combined with a triplet and metric-based few-shot learning approach to classify malware families with insufficient labeled samples. This framework addresses the challenge of obfuscation techniques by maintaining classification accuracy across different datasets~\cite{zhou2012dissecting}\cite{taheri2019extensible}. However, FAMCF's reliance on static features does not account for the sequential dependencies between API calls, potentially limiting its ability to fully capture the behavioral patterns of malware. This limitation may reduce its efficacy in scenarios where the sequence of actions is critical for identifying malicious behavior. Similarly, Bai et al.~\cite{bai2020unsuccessful} introduced a Siamese network-based method that improves the classification of few-shot Android malware families by embedding malware applications into a continuous vector space, enhancing accuracy across different family sizes. However, this approach primarily addresses data imbalance by incorporating both well-represented (large-scale) families and those with only a few samples (few-shot families) into the dataset, rather than focusing exclusively on the few-shot problem.

To the best of our knowledge, VOLTRON is the first study to propose a zero-shot learning approach for detecting Android malware. Additionally, by utilizing graphs that semantically represent code comprehensively, VOLTRON introduces a novel approach to enhance the effectiveness of malware detection. Its effectiveness and robustness against new malware families are thoroughly evaluated on one of the largest Android datasets.

\begin{figure*}[ht]
  \centering
  \includegraphics[width=\textwidth]{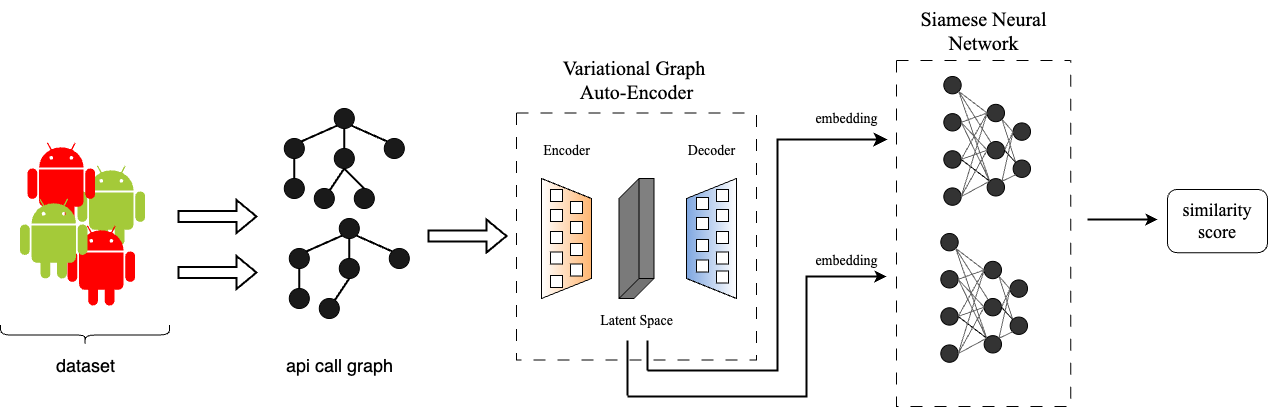}
  \caption{Overview of VOLTRON}
  \label{fig:generalSchema}
\end{figure*}

\section{Methodology}
\label{sec:methodology}

Our malware detection approach consists of two main components: the Variational Graph Auto-Encoder (VGAE)~\cite{kipf2016variational} and the Siamese Neural Network (SNN). These components are integrated to perform graph classification using graph embeddings extracted from Android applications. The conceptual schema of the proposed approach is illustrated in Fig.~\ref{fig:generalSchema}.

Firstly, applications are decompiled using Androguard~\cite{androguard}. Then, they are analyzed to extract features such as API calls and method interactions. These features are then used to construct API call graphs, which capture the complex relationships and control flow patterns within the application. The VGAE processes these graphs to learn low-dimensional representations. VGAE is particularly effective at capturing the essential structural features of the graphs, providing a latent space representation that is both detailed and succinct.

The SNN component takes the graph embeddings generated by the VGAE component and aims to learn the similarity between pairs of graph embeddings. SNN consists of two  identical subnetworks that employ the same architecture and weights. These subnetworks transform the input embeddings into a lower-dimensional space, enabling the network to focus on the most critical features. The absolute difference between the embeddings is then computed, capturing the dissimilarities between the two inputs. This difference is further processed to predict a similarity score, which indicates the likelihood that the two input graphs belong to the same class. More detailed information about each component and the training process is provided in the subsequent chapters. Finally, the details of how the proposed model is utilized for zero-shot learning are given.

\subsection{API Call Graph Construction}
\label{sec:graph-construction}

Firstly, the applications are decompiled into Smali code using Androguard \cite{androguard}. Then, instructions associated with common libraries (e.g., \texttt{Landroid}, \texttt{Ljava}, \texttt{Lcom/google}) that are less pertinent to the core analysis are filtered out. These operations, while essential for the app’s functioning, typically relate to standard Android or third-party libraries and do not significantly contribute to understanding the application's unique behavior. By focusing on the most relevant instructions, this process reduces the complexity of the analysis and eliminates extraneous information, enabling us to concentrate on code segments that more effectively represent the app's distinctive behavior.

Next, Axplorer~\cite{backes2016demystifying}, a comprehensive tool designed to map Android APIs and their associated permissions, is employed to identify critical APIs linked to dangerous permissions. These permissions grant applications access to sensitive or potentially exploitable features of the Android operating system. Since Axplorer covers mappings from API level 16 to API level 25, it encompasses the range of older applications in our dataset and is thus utilized in this study. 

Then, we extend the API list by including calls related to Java Security, Java Cryptography Extension (JCE), and Dalvik dynamic code loading libraries. These APIs are often utilized in advanced techniques employed by malware, such as encryption, code injection, or dynamic code execution. Including these additional APIs increased the total number of API calls under consideration from 2,743 to 3,121, providing a broad yet targeted set of features for our analysis. We subsequently extract a subset of these API calls (658) used in our training dataset, which includes both malicious and benign samples. This extraction reduces the size of the graphs we create for analysis. 

\begin{figure*}[ht]
  \centering
  \includegraphics[scale=0.40]{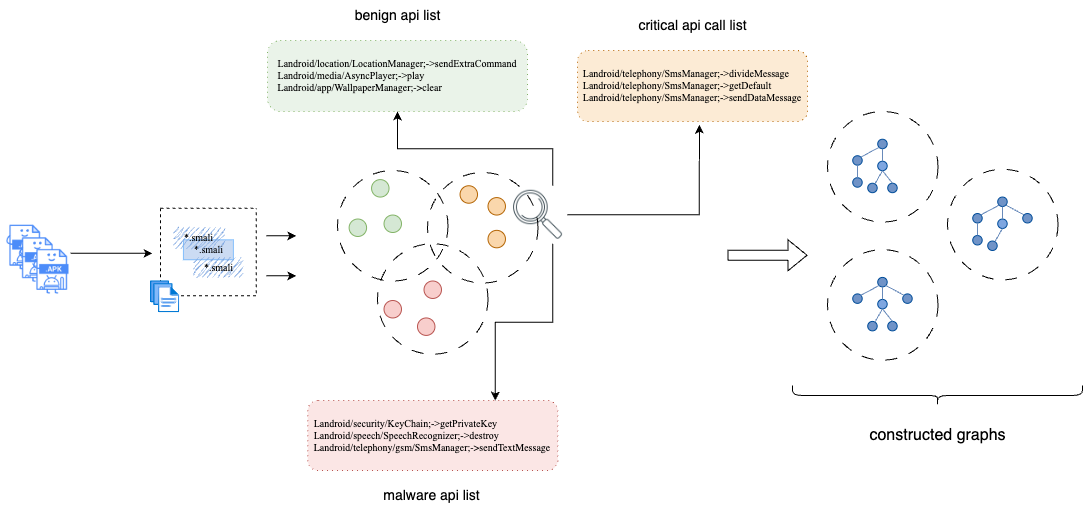}
  \caption{API Call Graph Construction}
  \label{fig:apiCallConstruction}
\end{figure*}

Finally, critical API call graphs are constructed, as illustrated in Fig.~\ref{fig:apiCallConstruction}, where nodes represent individual API calls and edges denote execution order relations derived from the code. API call graphs are first extracted for each method, after which they are merged based on calling relationships between methods, eliminating the need to specify entry points for graph construction.

\subsection{Variational Graph Auto-Encoder (VGAE)}
\label{sec:vgae}

The VGAE~\cite{kipf2016variational} is a deep learning model designed for graph-based tasks, including graph classification, where the goal is to predict the class label of an entire graph based on its structure and features. VGAE is based on the Variational Auto-Encoder (VAE)~\cite{rezende2014stochastic}\cite{kingma2013auto} framework and is designed to learn probabilistic latent representations that capture the essential features of graph-structured data. By leveraging latent variables, VGAE models uncertainty in the learned representations, which is particularly beneficial for complex graph structures.

In this study, VGAE is employed to learn low-dimensional representations of API call graphs for malware detection. VGAE’s capability to capture and model the underlying distribution of graph-structured data is crucial for distinguishing between malware and benign software. This is particularly important for identifying subtle and complex patterns within malware that may otherwise be obscured by noise or variability in the data. Notably, in the current representation, no explicit node or edge features are included, as the API calls forming the graph nodes do not possess intrinsic features. However, incorporating sensitive data flow as edge features in future work could enhance the detection of Android malware designed to exfiltrate sensitive data.

The VGAE model consists of two main components, as illustrated in Fig.~\ref{fig:generalSchema}: the encoder and the decoder. The encoder maps the input graph into a latent space, effectively capturing its structural properties. The decoder then reconstructs this latent representation, ensuring the preservation of essential features.

In the original VGAE framework, the reconstruction loss \( \mathcal{L}_{\text{rec}} \) ensures that the learned representations accurately capture the graph’s structure. It is extended by incorporating a regularization term into the final reconstruction loss:

\begin{equation}
\mathcal{L}_{\text{recon}} = \mathcal{L}_{\text{rec}} + \frac{1}{N} \mathcal{L}_{\text{KL}}
\end{equation}

where the Kullback-Leibler (KL) divergence loss (\( \mathcal{L}_{\text{KL}} \)) regularizes the latent space by encouraging the representations to follow a prior distribution. This regularization promotes well-behaved latent representations and mitigates overfitting, thereby improving generalization.

To further enhance the model, a classification component is introduced that utilizes the learned latent representations to predict the graph's class label. The classification loss is defined as:

\begin{equation}
\mathcal{L}_{\text{class}} = \text{CrossEntropy}(\text{logits}, \text{y})
\end{equation}

where \textit{logits} denote the predicted class scores, and \textit{y} represents the ground-truth labels. The overall loss function integrates these two components:

\begin{equation}
\mathcal{L} = \mathcal{L}_{\text{recon}} + \mathcal{L}_{\text{class}}
\end{equation}

This combined loss function allows the VGAE model to optimize for both reconstructing the graph's structure and accurately classifying the graph, enhancing its effectiveness for graph classification tasks.  By integrating these extensions, the model not only learns informative representations for reconstruction but also utilizes them for classification, thereby improving its versatility and performance.

The hyperparameters for VGAE were determined experimentally, considering the extended training times involved, rather than employing automated hyperparameter optimization techniques. These values are summarized in Table~\ref{tab:hyperparametersVGAE}.

\subsection{Siamese Neural Network (SNN)}
\label{sec:snn}

In this study, the SNN is employed to learn the similarity between pairs of embeddings, which represent the latent features of the input graphs. Unlike traditional neural networks that classify individual inputs independently, SNNs take pairs of embeddings and output a similarity score indicating whether the two embeddings likely belong to graphs of the same class. This makes SNNs particularly well-suited for tasks involving similarity measurement, such as face/signature verification, textual similarity, and anomaly detection. 

In the proposed approach, the SNN processes embeddings generated by the Variational Graph Auto-Encoder (VGAE). These embeddings, which encapsulate the learned latent features of the graphs, serve as inputs to the SNN, where they undergo further processing to determine their similarity.

Each SNN consists of two identical subnetworks, each responsible for processing one of the input embeddings derived from the VGAE. These subnetworks share the same architecture and weights, ensuring consistent treatment of both input embeddings and generating comparable representations. The process begins with each subnetwork receiving an embedding, which is passed through a series of fully connected layers. The first layer reduces the dimensionality of the input embedding, focusing the network on the most critical features by compressing the representation into a smaller vector. Subsequent layers further refine the embedding, with ReLU activation functions introducing non-linearity, essential for modeling complex relationships within the data.

Once the embeddings from both inputs have been processed through their respective subnetworks, the network calculates the absolute difference between the two resulting embeddings~\cite{koch2015siamese}. This step measures the degree of dissimilarity between the embeddings, capturing the differences in the features learned by the subnetworks. By focusing on these differences, the network can effectively assess whether the two inputs belong to the same class or not, thereby enabling accurate classification based on their similarities and dissimilarities.

The computed difference is then passed through additional fully connected layers, where the network processes the information to predict a similarity score. The final layer of the SNN applies a sigmoid activation function, outputting a probability score. This score indicates the likelihood that the two input embeddings, and thus the two graphs they represent, belong to the same class.

By comparing the similarity scores of various graph pairs, the SNN can effectively classify relationships, determining whether the embeddings correspond to benign or malicious graphs. This approach enables the model to identify and classify the relationships between graphs based on their structural features.

The hyperparameters used in the SNN, including the number of layers, the dimensionality of each layer, learning rate, and batch size, were carefully selected through an automated hyperparameter optimization process using Optuna~\cite{optuna_2019}. Optuna is a framework that efficiently optimizes hyperparameters by exploring a large search space with techniques such as Bayesian optimization. By leveraging Optuna with the default sampler, we identified the optimal set of hyperparameters that maximized the performance of our SNN for similarity measurement. The details of the selected hyperparameters are provided in Table~\ref{tab:hyperparameters}.

\begin{figure}[ht]
  \centering
  \includegraphics[width=\linewidth]{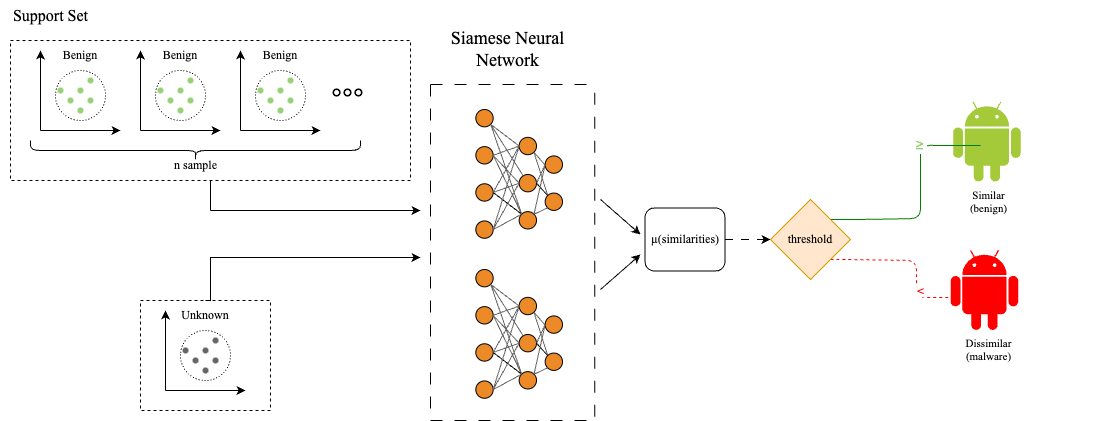}
  \caption{Zero-Shot Learning}
  \label{fig:zeroShot}
\end{figure}

\subsection{Zero-Shot Learning (ZSL)}
\label{sec:zsl}

In machine learning, particularly in the context of malware detection, the concepts of few-shot and zero-shot learning have emerged as vital approaches to overcome the limitations posed by traditional supervised methods. Few-shot learning (FSL) is a paradigm designed to train models with a very limited number of labeled samples, a significant advantage when dealing with novel or underrepresented malware families. Within FSL, zero-shot learning (ZSL) is a subset where the model is trained to classify instances of classes that were not present in the training set. In other words, while few-shot learning may have a small number of labeled samples from every class, zero-shot learning operates with no training samples from certain classes, making it particularly challenging and applicable to real-world scenarios where new, previously unseen threats emerge. This approach is essential in scenarios where traditional deep learning models, which rely heavily on large-scale labeled datasets, struggle to adapt to the continuous emergence of novel malware families.

In this study, the concept of zero-shot learning is applied to Android malware detection using a Variational Graph Auto-Encoder (VGAE) combined with a Siamese Neural Network (SNN). Unlike traditional machine learning setups that typically involve only a training set and a test set, our model introduces a third set known as the support set. The support set is crucial in the SNN's comparison process, where the unknown APK is compared against other applications. Here, the support set is exclusively composed of benign applications. The similarity between the unknown APK and the benign applications is measured, which allows us to determine whether the unknown APK is benign or malicious. By focusing on the similarity to benign samples, our model aims to detect malware based on its dissimilarity to benign behaviors, which is the core idea behind zero-shot malware detection. The proposed approach is summarized in Fig.~\ref{fig:zeroShot}.

An essential consideration in this approach is the composition of the support set. Evaluating the unknown APK against only a single benign sample could yield unreliable results due to the variability in benign applications, leading to false classifications. Therefore, our methodology includes comparing the unknown APK against multiple benign samples and averaging the similarity scores to achieve a more accurate assessment. After extensive testing with different support set sizes, we observed no significant impact on the results. Consequently, we opted for a support set size of 30 benign samples, selected randomly for each APK, to maintain consistency and efficiency.

Regarding the threshold selection, we experimented with various threshold values and found that accuracy remained stable between 0.4 and 0.5. To optimize for the best performance, we selected 0.5 as our final threshold, as it yielded the highest F1-score. If the calculated similarity score exceeds this threshold, the APK is classified as benign; otherwise, it is deemed malicious. The threshold selection process is illustrated in \ref{fig:accuracy_f1_threshold}.

\section{Evaluation} \label{sec:evaluation}

In this section, we first introduce the dataset used for training and testing, followed by a description of the experimental settings. We then present and discuss the experimental results, concluding with a comparison of our approach to the state of the art and few-shot learning.

\subsection{Dataset}

In this study, we used the KronoDroid dataset~\cite{guerra2021kronodroid}, a comprehensive and diverse collection of Android benign and malicious software samples. It incorporates samples from various sources, including Drebin~\cite{arp2014drebin}, AMD~\cite{wei2017deep}, VirusTotal~\cite{virustotal2020}, VirusShare~\cite{virusshare2020}, F-Droid~\cite{fdroid2020}, MARVIN~\cite{lindorfer2015marvin}, and APKMirror~\cite{apkmirror2020}, spanning different malware families and time periods. Comprising approximately 60,000 samples collected between 2008 and 2020, KronoDroid is one of the largest labeled datasets in the literature.

Samples in this dataset were classified based on their malware detection ratios and the legitimacy of their sources. Samples were labeled as benign if they had a malware detection ratio of zero and originated from a trusted, legitimate source. Conversely, samples were labeled as malware if they had a non-zero detection ratio, were identified in a recognized malware repository, and were associated with a specific malware family. In addition to malicious and benign samples, the dataset includes samples labeled as indefinite, which do not fit into either of these categories. Consequently, these indefinite samples were removed from our dataset. Additionally, samples that were mistakenly classified as both malware and benign in the original KronoDroid dataset were filtered out. To further clean the dataset, we addressed the issue of redundancy. Since the package name serves as the primary identifier for each APK, we retained only one version of each APK with the same package name. This step was essential to avoid potential bias that could arise from including multiple versions of the same application.
\vspace{-5mm}
\begin{figure}[ht]
\centering
\includegraphics[width=0.65\columnwidth, height=0.6\textheight, keepaspectratio]{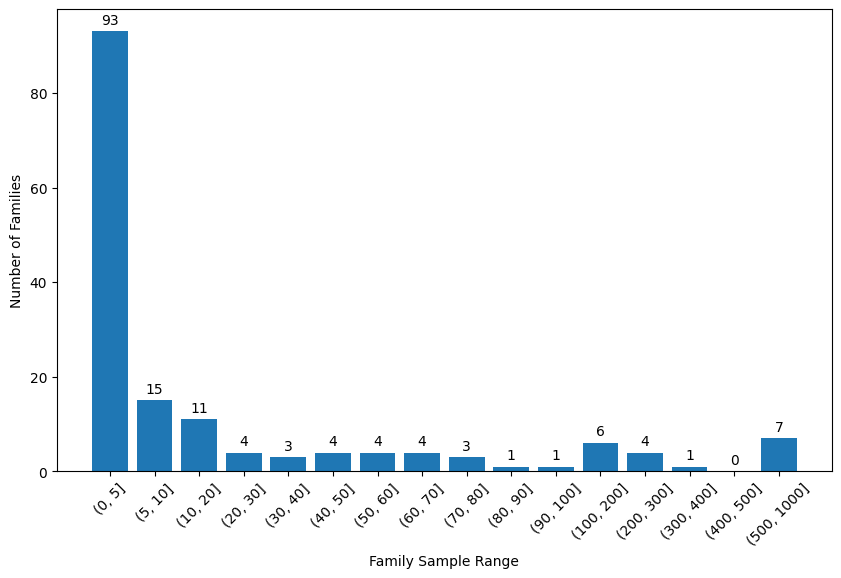}
\caption{Distribution of Sample By Malware Family}
\vspace{-5mm}
\label{fig:sampleDist}
\end{figure}

Finally, we excluded any applications that failed to produce valid graphs during the API call graph construction phase due to some errors such as invalid instruction errors. After completing these pre-processing steps, our final dataset consisted of 28,294 malware samples from 218 distinct families and 27,713 benign applications. The details of the dataset are summarized in Table~\ref{tab:Dataset}.

We also collected 400 APKs for each of the years 2021 and 2022 from AndroZoo~\cite{androzoo}, each year comprising 200 benign and 200 malware samples. These newer samples were used exclusively in our time-based experiments to assess how well our approach detects threats emerging after the training period.

\begin{table}[h!]
\centering
\caption{The Dataset}
\label{tab:Dataset}
\begin{tabularx}{\textwidth}{@{}lXXXX@{}}
\toprule
\renewcommand{\arraystretch}{0.8}
\textbf{Set} & \textbf{Malware} & \textbf{Family} & \textbf{Benign} & \textbf{Total} \\
\midrule
Train   & 23,065 & 164 & 21,290 & 44,355 \\
Test    &  5,229 &  54 &  5,323 & 10,552 \\
Support &      0 &   0 &    100 &    100 \\
\midrule
Total   & 28,294 & 218 & 26,713 & 55,007 \\
\bottomrule
\end{tabularx}
\end{table}

\subsection{Experimental Settings}

Firstly, we evaluate our model's ability to detect previously unknown malware families by using zero-shot learning. Our goal here is to simulate real-world scenarios, such as zero-day environments, by introducing malicious samples into the test set that belong to families not included in the training set. This approach contrasts with traditional machine learning methods, which typically evaluate models using randomly split training and test sets, allowing the same malware families to appear in both. By excluding entire families from the training set, our method offers a more realistic evaluation of how the model would perform against new, previously unseen threats.

In this study, we apply a family-based split with an 80-20 ratio. Here, any malware family present in the test set was excluded from the training set. For instance, if the dataset consisted of malware families \(\{A, B, C, D, E\}\) and benign samples, the training set \(T\) included \(\{A, B, C\}\), while the test set included \(\{D, E\}\). Additionally, we incorporated a support set composed of \(n\) benign samples that were not part of either the training or test sets. The main goal is to evaluate whether the model can accurately detect malware from families that it has not encountered during training.

The proposed approach is compared with MaMaDroid~\cite{onwuzurike2019mamadroid}, which is well-regarded and widely recognized in the field of Android malware detection. Additionally, it has demonstrated the ability to detect previously unknown malware over time, thereby reducing the need for continuous retraining. To ensure a fair comparison, we replicated the experimental setup of MaMaDroid and trained it using our training set. We then evaluated its performance with the same test set given in Table~\ref{tab:Dataset}. Furthermore, the comparison is conducted across different train-test splits of the dataset to demonstrate the robustness and consistency of both methods. The number of families in each fold is demonstrated in Table~\ref{tab:cross_validation_family_counts}. Notably, the families in each fold are distinct.

The proposed approach is also compared with few-shot learning approach. In this method, the support set includes both benign samples and samples from malware families present in the test set. This approach allowed us to simulate real-world scenarios where only a limited number of samples are available for new malware families. In this few-shot setup, the support set contained both malware and benign samples, enabling the model to compare the unknown APK against both types during classification. Unlike the zero-shot experiments, where classification was based on a fixed threshold value, the few-shot experiment involved comparing the average probabilities of the unknown sample being classified as malware or benign. This approach provided a more nuanced understanding of the model’s classification capabilities in practical contexts. While 30 benign samples are utilized in the zero-shot setting, the few-shot setup incorporates a balanced support set consisting of 30 benign and 30 malware samples.

In addition to the standard evaluation splits, we conducted a time-based experiment to measure the model’s resilience to new threats that appear after the training period. Specifically, we randomly collected 400 APKs for each of the years 2021 and 2022—comprising 200 benign and 200 malicious samples per year—from the AndroZoo repository \cite{androzoo}. We trained our model on KronoDroid samples spanning up to 2020 and then tested it on these newly collected APKs. This approach simulates a realistic scenario where unknown malware families emerge well beyond the training window, allowing us to assess how effectively the model adapts to evolving malicious behaviors and any changes in the Android API over time. The proposed approach is also evaluated on various obfuscation techniques, namely name obfuscation, variable encryption, API reflection and string encryption, to assess its resilience against code obfuscation strategies commonly employed by malware developers. Lastly, an ablation study is conducted to analyze the impact of each model component on overall performance

\subsection{Experimental Results}

\subsubsection{The Performance of VOLTRON}

We first demonstrate the performance of our zero-shot approach in detecting new malware families in Table~\ref{tab:model-comparison}. As shown, the proposed method achieves a detection rate of 95.20\% for new families with an acceptable false positive rate (FPR) of 2.72\%. Additionally, we compare our approach with MaMaDroid~\cite{onwuzurike2019mamadroid}, which employs different classification algorithms, including Random Forests (RF), 1-Nearest Neighbor (1-NN), and 3-Nearest Neighbor (3-NN). We have included comparisons with the performance of each of these classifiers. As reported in~\cite{onwuzurike2019mamadroid}, the best performance for MaMaDroid is achieved using Random Forests (RF). Similarly, in our experiments, we also obtained the best results with RF. 

\begin{table}
\centering
\caption{Evaluation of VOLTRON and Comparison with MaMaDroid~\cite{onwuzurike2019mamadroid}}
\label{tab:model-comparison}
\resizebox{\textwidth}{!}{
\begin{tabular}{lccccc}
\toprule
\textbf{Model} & \textbf{Accuracy (\%)} &\textbf{ Recall (\%)} & \textbf{Precision (\%)} & \textbf{F1-score (\%)} &\textbf{ FPR (\%) }\\
\midrule
VOLTRON & 96.24 & 95.20 & 97.17 & 96.17 & 2.72 \\
MaMaDroid (RF) & 91.91 & 89.11 & 95.20 & 92.05 & 2.50 \\
MaMaDroid (3-NN) & 89.61 & 88.32 & 91.63 & 89.94 & 4.52 \\
MaMaDroid (1-NN) & 87.73 & 85.90 & 90.30 & 88.05 & 5.24 \\
\bottomrule
\end{tabular}
}
\vspace{-5mm}
\end{table}

The comparison shows that our method achieves a significantly higher detection rate of 95.20\% for new malware families, compared to MaMaDroid's 89.11\%. Importantly, this improvement in detection capabilities comes with only a slight increase in the false positive rate, which is 2.72\% for our method compared to MaMaDroid's 2.5\%. This demonstrates that our model not only detects malware more effectively but also reliably distinguishes between malicious and benign samples. This underscores the robustness and precision of our approach, particularly in scenarios involving previously unseen malware families.

The time-based evaluation results \ref{tab:timebased} indicate that while the model remains effective, its performance shifts when tested on newer malware from 2021-2022. The model maintains a strong recall (85.00\%), but its accuracy declines from 96.24\% to 82.25\%, and the false positive rate increases to 20.00\%. This suggests that malware and benign app behaviors are evolving as expected. 
These findings emphasize the necessity of regular retraining with updated datasets to sustain high detection rates and minimize false positives, ensuring continued robustness against emerging threats.

\begin{table}
    \centering
    \caption{Time-Based Evaluation of the Approach}
    \label{tab:timebased}
    \renewcommand{\arraystretch}{0.7}
    \begin{tabular}{lccccc}
        \toprule
        \textbf{Setting} & \textbf{Accuracy} & \textbf{Precision} & \textbf{Recall} & \textbf{F1-score} & \textbf{FPR} \\
        \midrule
        2012-2020 & 96.24 & 97.17 & 95.20 & 96.17 & 2.72 \\
        2021-2022 & 82.25 & 80.31 & 85.00 & 82.97 & 20.00 \\
        \bottomrule
    \end{tabular}
\end{table}

The results for all families are demonstrated in Figure~\ref{fig:family-detection-rates}. The number of test samples for each family is also indicated on the corresponding bars in the bar chart. As illustrated in Figure~\ref{fig:family-detection-rates}, certain families exhibit lower detection rates, and 7 families out of 54 test families are not detected at all. It is worth noting that these families have only a few samples in the dataset. Upon analyzing these families, we identified several reasons for these shortcomings. Four of these families, namely ClickFraud, Downloader, Sakezon, and PDAspy, exhibit behaviors that lie in a grey area between benign and malicious, which can lead to misclassification by detection algorithms that depend on well-defined boundaries between benign and malicious behaviors. 

A notable example is PDAspy, which is a commercial monitoring application that logs phone calls, SMS messages, and GPS locations. While PDAspy can have legitimate use cases such as parental monitoring, it is also prone to misuse, making it challenging for models to classify accurately. This dual-use nature often results in PDAspy being flagged by security solutions as riskware~\cite{fortiguard}\cite{fsecure}, yet the inherent complexity of such applications can obscure the classification process, reducing the accuracy of machine learning models that are designed to detect more straightforward instances of malware.

To further evaluate the performance and generalization ability of our approach, we implemented five-fold cross-validation by testing it on multiple subsets of the dataset. Specifically, we divided the families into five parts, ensuring that the number of examples in each part was as balanced as possible. In each experiment, one part was used for testing, while the remaining four were used for training, maintaining an approximate 80-20 ratio. We also ensured that no family present in the training set appeared in the test set. This approach enabled us to assess the model's performance across different data subsets, minimizing the risk of bias from any single train-test split. The objective was to provide a more accurate evaluation of the model’s generalization capabilities, particularly its ability to perform well on unseen malware families, resulting in a more reliable overall assessment. The same API calls extracted from the training set in Table~\ref{tab:Dataset} is also used to generate API call graphs in this experiment.
\begin{table}
    \centering
    \caption{Model performance across 5 folds}
    \renewcommand{\arraystretch}{0.6}
    \label{tab:folds}
    \resizebox{\textwidth}{!}{%
        \begin{tabular}{lcccccc}
            \toprule
            \textbf{Model} & \textbf{Fold} &   \textbf{Accuracy (\%)} & \textbf{Precision (\%)} & \textbf{Recall (\%)} & \textbf{F1 (\%)} & \textbf{FPR (\%)} \\ 
            \midrule
            MaMaDroid  & 1 & 67.61 & 93.94 & 47.80 & 63.36 & 2.60 \\ 
            VOLTRON  &  & 77.78 & 93.64 & 65.89 & 77.70 & 2.90 \\ 
            \midrule
            MaMaDroid & 2  & 92.69 & 94.01 & 91.22 & 92.59 & 3.05 \\ 
            VOLTRON &  & 95.12 & 94.31 & 92.09 & 93.20 & 3.83 \\ 
            \midrule
            MaMaDroid & 3  & 91.63 & 93.58 & 89.40 & 91.44 & 3.24 \\ 
            VOLTRON  &  & 95.61 & 96.42 & 94.38 & 95.39 & 4.91 \\ 
            \midrule
            MaMaDroid  & 4    & 84.99 & 94.26 & 74.48 & 83.21 & 2.60 \\ 
            VOLTRON  &   & 88.36 & 96.09 & 83.00 & 89.13 & 3.12 \\ 
            \midrule
            MaMaDroid  & 5    & 80.69 & 93.51 & 65.88 & 77.50 & 2.75 \\ 
            VOLTRON &    & 93.09 & 96.68 & 89.81 & 93.12 & 3.34 \\
            \midrule
            MaMaDroid & Avg.    & 83.52 & 93.86 & 73.36 & 81.62 & 2.85 \\ 
            VOLTRON  & Avg.  & 89.73 & 95.42 & 84.96 & 89.66 & 3.86 \\
            \bottomrule
        \end{tabular}%
    }
\end{table}
\subsubsection{Comparison with MaMaDroid}

The results of each fold are compared with MaMaDroid in Table \ref{tab:folds}. Our model consistently outperforms MaMaDroid's approaches 
across all folds. 
It indicates our model's effectiveness in identifying malware instances. The cross-validation results also highlight the reliability of our model in maintaining a balanced performance between precision and recall while keeping the false positive rate low. This ensures that it accurately distinguishes between malware and benign applications. In contrast, MaMaDroid showed more variability across different folds, with lower recall and F1 scores. This suggests that our model offers a more stable and reliable solution for detecting malware in diverse data environments, reinforcing its suitability for real-world applications.

While the proposed approach provides a high detection rate (recall) across all folds, the first fold of the cross-validation shows lower performance compared to the others, which can be attributed to the composition of the test dataset as demonstrated in Table \ref{tab:cross_validation_family_counts}. Specifically, this fold contains only samples from the Airpush/StopSMS family, which is typically classified as adware. Adware usually displays advertisements automatically and, though often not malicious, it is generally considered unwanted by users. This dual nature of adware can make it challenging for detection models, as adware may not exhibit the same clear-cut malicious behaviors as traditional malware. This is particularly relevant for the Airpush family, which is known for aggressive advertising practices but does not always engage in malicious activities, potentially leading to lower recall rates during classification.

\subsubsection{Zero-shot vs. Few-Shot Learning}

The performance of zero-shot/few-shot learning is illustrated in Table \ref{tab:resultsfewshot}. While the few-shot setup achieved a test accuracy close to that of the zero-shot model. However, the false positive rate (FPR) improved with the inclusion of additional malware samples in the support set.

\begin{figure}
    \centering
    \includegraphics[width=\textwidth]{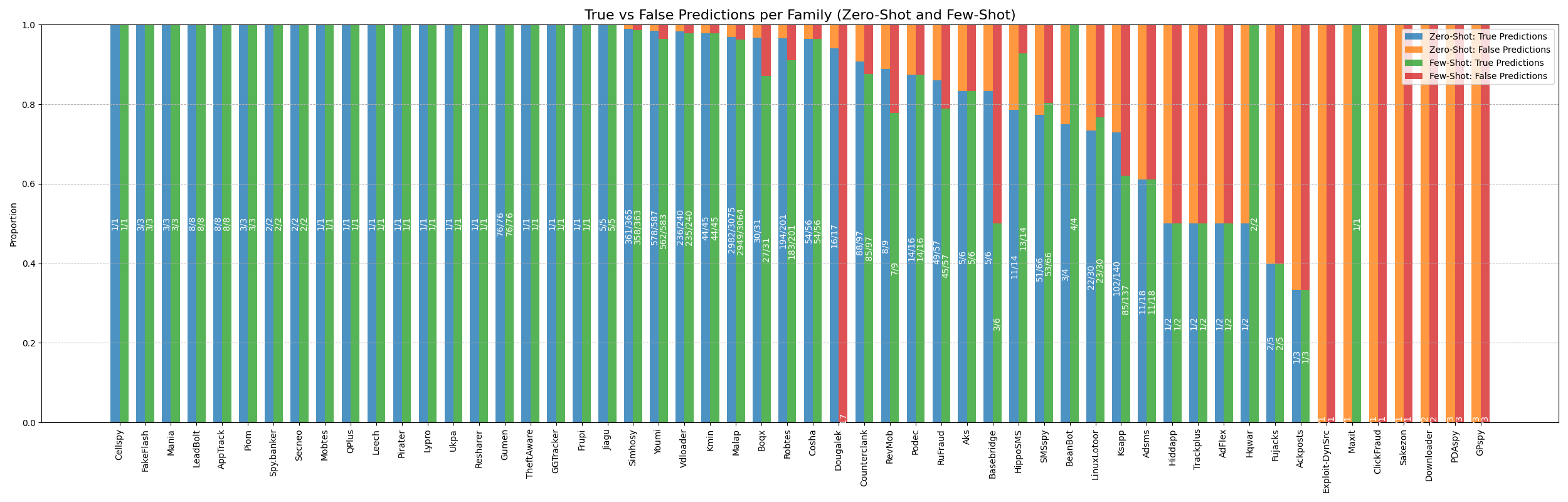}
    \caption{Zero-Shot vs. Few-Shot in Detecting New Malware Families}
    \vspace{-10mm}
    \label{fig:fewshot_family-detection-rates}
\end{figure}

Since zero-shot learning has already achieved very high detection rates, few-shot learning does not demonstrate any improvement. Additionally, in the few-shot setting, the support set consists of only 30 malware families randomly selected from the test set to ensure a balanced comparison with zero-shot learning, which does not encompass all families. Hence, if the few-shot samples lack sufficient diversity, they may not adequately represent the variations within a family. On the other hand, the positive impact of few-shot learning is observed in the false positive rate. Including malware samples in the support set enhances the model's ability to differentiate between benign and malicious applications, leading to improved accuracy in detecting true negatives. The results for all families are demonstrated in Figure \ref{fig:fewshot_family-detection-rates}.

\begin{table}
    \centering
    \caption{Zero-Shot vs. Few-Shot}
    \label{tab:resultsfewshot}
    \renewcommand{\arraystretch}{0.7}
    \begin{tabular}{lccccc}
        \toprule
        \textbf{Setting} & \textbf{Accuracy} & \textbf{Precision} & \textbf{Recall} & \textbf{F1-score} & \textbf{FPR} \\
        \midrule
        Zero-Shot & 96.24 & 97.17 & 95.20 & 96.17 & 2.720 \\
        Few-Shot & 96.30 & 97.22 & 95.24 & 96.22 & 2.667 \\
        \bottomrule
    \end{tabular}
\end{table}

\subsubsection{Robustness to Obfuscation Techniques.}

In this experiment, we assess our model's resilience against obfuscation techniques. We modified our applications using AVPass~\cite{jeon2017avpass} and implemented specific methods, including name obfuscation, variable encryption, API reflection, and string encryption. Name obfuscation modifies class and method names to hinder code analysis, making it more challenging to understand the underlying logic. Variable encryption involves encrypting the values of variables within the code, hiding their true content until they are decrypted at runtime. Similarly, string encryption conceals hardcoded strings, such as URLs or commands, decrypting them only when needed during execution to evade detection. API reflection is used to dynamically invoke Android APIs at runtime instead of referencing them directly, making it difficult for static analysis tools to detect malicious API calls. 

The objective of this experiment is to evaluate the model's ability to maintain detection accuracy against evasion techniques employed by malware authors, demonstrating its robustness and adaptability in scenarios where malware is often obfuscated to evade detection.

The results are presented in Table~\ref{tab:obfuscation_results}. While all obfuscation techniques lead to a decline in detection rates, string encryption and API reflection have the most significant impact. The reduction in performance due to string encryption can be attributed to the intrinsic challenges associated with this technique. In string encryption, critical strings—such as URLs, API names, and system services—are encrypted, making it difficult for static analysis to accurately identify and map essential API calls. Consequently, the model's reliance on API call graphs is severely undermined; the inability to accurately map these encrypted strings results in incomplete or inaccurate graph structures, ultimately reducing the model's effectiveness in detecting malicious patterns.

\begin{table}[htbp]
    \centering
    \caption{Performance Evaluation Against Obfuscation Techniques}
    \label{tab:obfuscation_results}
    \resizebox{\columnwidth}{!}{%
        \begin{tabular}{lcccc}
            \toprule
            \textbf{Obfuscation Type} & \textbf{Accuracy (\%)} & \textbf{Precision (\%)} & \textbf{Recall (\%)} & \textbf{F1-score (\%)} \\
            \midrule
            Baseline  & 96.24 & 97.17 & 95.20 & 96.17 \\
            Name Obfuscation & 93.68 & 95.95 & 91.09 & 93.46 \\
            Variable Encryption & 96.08 & 97.18 & 94.84 & 95.99 \\
            String Encryption & 77.16 & 91.23 & 59.65 & 72.13 \\
            API Reflection  & 51.90 & 72.05 & 4.78 & 8.97 \\
            \bottomrule
        \end{tabular}
    }
\vspace{2mm}
\end{table}

The results for API reflection obfuscation revealed substantial challenges. API reflection enables malware to dynamically invoke methods, effectively bypassing static API analysis. This dynamic invocation obscures the actual methods called during execution, resulting in incomplete or misleading call graphs. Such obfuscation complicates both static and dynamic analysis efforts, as the true flow of execution remains hidden from the model. The decline in detection capability under API reflection underscores the difficulty of identifying malware when critical behaviors are masked at runtime. Future research should explore advanced static-dynamic hybrid analysis, de-obfuscation strategies~\cite{bichsel2016statistical} to more effectively address the complexities introduced by these obfuscation techniques.

\subsubsection{Ablation Study}

Lastly, we conducted an ablation study to evaluate the impact of the zero-shot learning technique on malware detection. In this experiment, we utilized embeddings generated by VGAE to classify applications. We first processed these embeddings through a global mean pooling operation to obtain a pooled representation. This representation was then fed into a linear classifier, which outputs a probability distribution over the class labels to determine whether an application is classified as malware or benign.

Table \ref{tab:ablation} shows the performance metrics of VGAE (without zero-shot learning) compared to VOLTRON. The results show that VOLTRON achieved a higher accuracy (96.24\%) compared to VGAE without zero-shot learning (94.83\%), indicating an overall improvement in the model’s correctness. The inclusion of zero-shot learning notably enhanced the detection rate (3.84\%) for new malware families. However, this improvement comes with an approximate 1\% increase in the false positive rate. It is worth noting that the graph-based malware detection system using VGAE (without ZSL) continues to outperform MaMaDroid both in terms of detection rate and false positive rate, highlighting the advantages of representing code as graphs for malware detection. This supports our hypothesis that graphs are a natural and semantically meaningful representation of code.

\begin{table}
\centering
\scriptsize
\caption{Ablation Study}
\label{tab:ablation}
\begin{tabular}{@{}lccccc@{}}
\toprule
\textbf{Model} & \textbf{Accuracy (\%)} & \textbf{Recall (\%)} & \textbf{Precision (\%)} & \textbf{F1-score (\%)} & \textbf{FPR (\%)} \\
\midrule
VGAE (without 0-shot) & 94.83 & 91.36 & 98.09 & 94.60 & 1.75 \\
VOLTRON & 96.24 & 95.20 & 97.17 & 96.17 & 2.72 \\
MaMaDroid (RF) & 91.91 & 89.11 & 95.20 & 92.05 & 2.50 \\
\bottomrule
\end{tabular}
\end{table}

In summary, the results underscore the effectiveness of our proposed model in malware detection. Not only does it surpass the performance of established methods like MaMaDroid, but it also shows a strong ability to generalize to newmalware families and to operate effectively with limited training data. These strengths make our model a promising candidate for deployment in real-world malware detection systems, where adaptability and accuracy are paramount.

\section{Limitations}
\label{sec:limitations}

\color{black}
This is the first study that aims to detect unknown malware using zero-shot learning techniques. While this approach represents a significant advancement, it is important to acknowledge its limitations.

The KronoDroid dataset, despite being one of the largest available in terms of malware family size, presents a significant challenge due to the uneven representation of different malware families, as shown in Figure \ref{fig:sampleDist}. The substantial disparity in sample sizes across these families in the training set can impact the model's performance, as shown in Table \ref{tab:folds}. During training, underrepresented classes may lead to insufficient learning, which can cause the model to struggle with generalizing effectively to these less-represented or rare malware families during testing. This imbalance may particularly affect the model’s performance when encountering malware families that are either very rare or disproportionately common. Therefore, it is essential to provide a well-represented dataset of real malware samples and include a broader range of known malware families. 

Furthermore, we encountered challenges in accurately identifying malware family names due to inconsistencies across various sources. Different antivirus vendors and security researchers often use distinct naming conventions for the same malware family, and these names can evolve over time. For instance, a malware family identified under one name by a particular security organization today may be reclassified under a different name by the same or another organization in the future. Such inconsistencies can complicate the process of labeling unknown families and subsequently affect the evaluation of our model's performance in detecting these families.

Another key limitation is the temporal variability of Android APIs, including changes in API levels, the introduction of new APIs, modifications to existing APIs, and the deprecation of older ones. Such time-specific features can impact the performance of classifiers \cite{pendlebury2019tesseract,liu2022explainable}, as current API call graphs may not fully capture the ongoing changes in the Android ecosystem. Consequently, the system must be regularly updated and retrained to maintain its effectiveness.

Moreover, certain obfuscation techniques, particularly API reflection, challenge our model by disrupting the structure of API call graphs. These methods hinder accurate extraction and analysis of application behavior, potentially reducing the effectiveness of our approach. Addressing this issue will require additional strategies, such as incorporating dynamic analysis to capture behaviors at runtime, thereby managing the complexities introduced by these obfuscation techniques. This highlights an important area for future improvement.

Lastly, the construction of sensitive API call graphs, which is fundamental to our approach, relies heavily on the accuracy of permission-API call mappings provided by tools such as Axplorer \cite{backes2016demystifying} or Arcade \cite{aafer2018precise}. The reliability of these mappings is crucial, as any inconsistencies could result in questionable intersections or missing data in the final analysis, potentially affecting the overall performance of the model. Additionally, our selection of sensitive API calls for constructing API call graphs depends not only on the mappings provided by Axplorer but also on the intersection of benign and malware API calls within our dataset. Consequently, the specific set of API calls used in our analysis may vary with different datasets or tools.

\section{Conclusions}
\label{sec:conclusion}

In this study, we proposed a novel approach to Android malware detection using zero-shot learning, combining Variational Graph Auto-Encoders (VGAE) and Siamese Neural Networks (SNN). Our method was designed to tackle the challenge of detecting previously unseen malware families, a critical issue in the rapidly evolving landscape of cybersecurity. By leveraging graph-based representations of Android applications, VOLTRON effectively captured the structural relationships within the data, allowing for robust classification even with no prior examples of certain malware families.

The experimental results demonstrated that our approach outperformed the existing method, MaMaDroid, in both accuracy and detection rates, particularly in zero-shot scenarios. These findings suggest that zero-shot learning, as implemented in our framework, can significantly enhance future malware detection systems by improving protection against emerging threats without the need for extensive retraining or large, labeled datasets.

\bibliographystyle{unsrt}  
\bibliography{references}

\appendix
\section{Appendix}
\begin{figure}[ht]
\centering
\includegraphics[width=0.6\columnwidth, height=0.6\textheight, keepaspectratio]{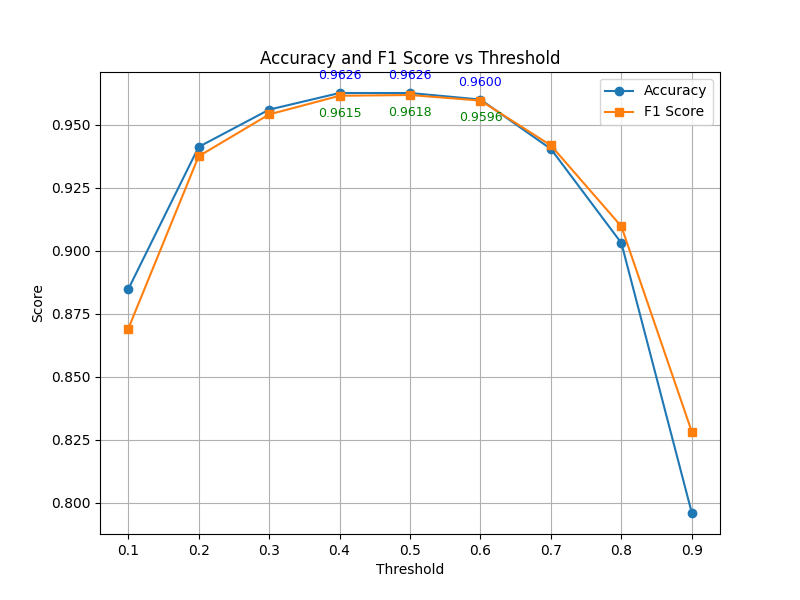}
\caption{Accuracy and F1 Score as a Function of the Classification Threshold}
\label{fig:accuracy_f1_threshold}
\end{figure}

\begin{figure}
    \centering
    \includegraphics[width=\linewidth]{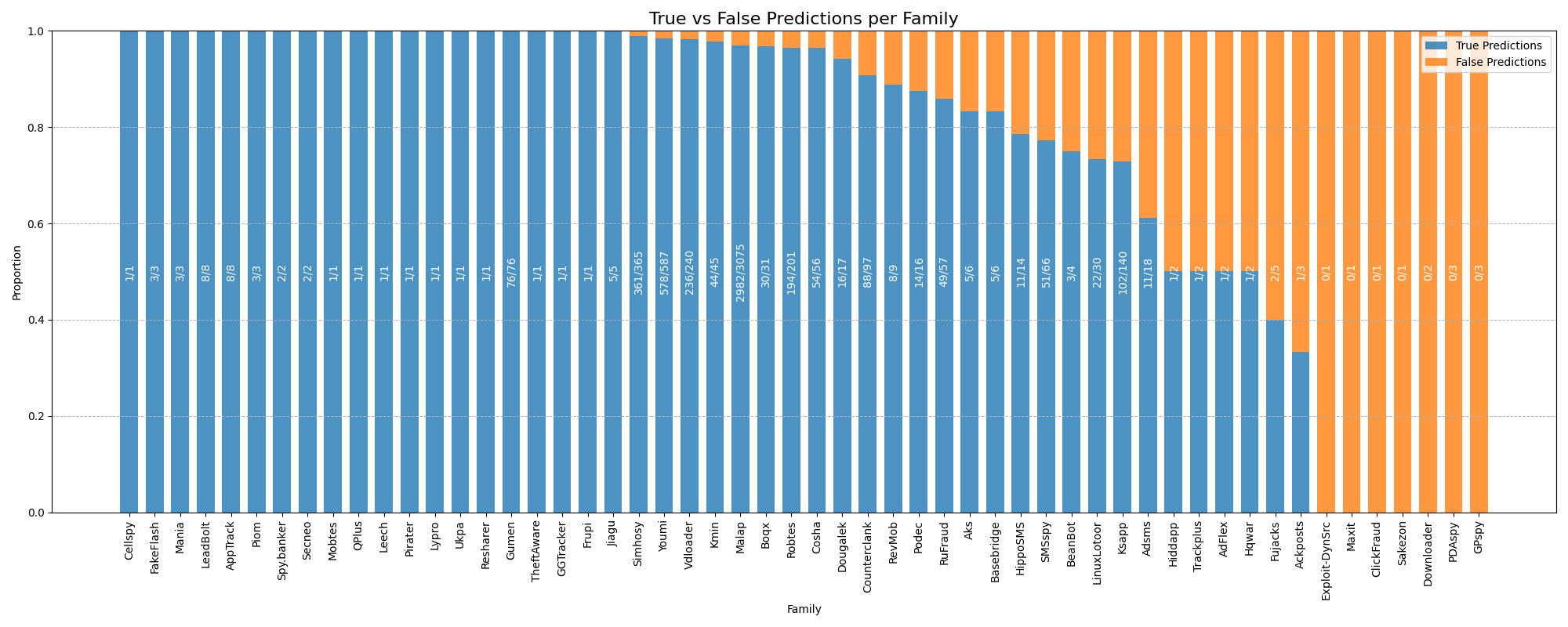}
    \caption{Performance of VOLTRON in Detecting New Malware Families}
    \label{fig:family-detection-rates}
\end{figure}

\begin{table}[htbp]
\centering
\caption{Hyperparameters of VGAE}
\small
\begin{tabular}{ll}
\toprule
\textbf{Parameter} & \textbf{Value} \\
\midrule
Learning Rate      & 0.001 \\
Number of Epochs   & 300 \\
Hidden Channels    & [32, 24] \\
Optimizer         & Adam \\
Latent Dimension  & 16 \\
\bottomrule
\end{tabular}
\label{tab:hyperparametersVGAE}
\end{table}

\begin{table}[htbp]
\centering
\caption{Hyperparameters of SNN}
\small 
\begin{tabular}{ll} 
\toprule
\textbf{Parameter} & \textbf{Value} \\
\midrule
Learning Rate         & 0.001 \\
Number of Epochs      & 4 \\
Optimizer            & SGD \\
Support Set Size      & 30 \\
Hidden Layer Dimensions & [128, 64, 32] \\
\bottomrule
\end{tabular}
\label{tab:hyperparameters}
\end{table}

\begin{table}
    \centering
    \caption{Number of Families in Each Fold}
    \label{tab:cross_validation_family_counts}
    \small
    \renewcommand{\arraystretch}{0.8}
    \begin{tabular}{lcc} 
        \toprule
        \textbf{Fold} & \textbf{Train Families} & \textbf{Test Families} \\ 
        \midrule
        1 & 217 & 1  \\
        2 & 168 & 50  \\
        3 & 164 & 54  \\
        4 & 162 & 56  \\
        5 & 161 & 57 \\
        \bottomrule
    \end{tabular}
\vspace{-5mm}
\end{table}

\end{document}